\documentclass[aps,prl,reprint,groupedaddress]{revtex4-2}

\usepackage{amsmath,graphicx,color}
\usepackage{bm}

\newcommand{\pd}{{\phantom{\dagger}}}

\newcommand{\FST}{FeSe$_{0.45}$Te$_{0.55}$ }
\newcommand{\FSTT}{FeSe$_{0.45}$Te$_{0.55}$}

\begin{document}

\title{Origin of Topological Surface Superconductivity in \FST}

\author{Eric Mascot$^{1}$, Sagen Cocklin$^{1}$, Martin Graham$^{1}$, Mahdi Mashkoori$^{2}$, Stephan Rachel$^{2}$, and Dirk K. Morr$^{1}$}
\affiliation{$^{1}$University of Illinois at Chicago, Chicago, IL 60607, USA}
\affiliation{$^{2}$School of Physics, University of Melbourne, Parkville, VIC 3010, Australia\\}

\date{\today}


\begin{abstract}

\end{abstract}

\maketitle

{\bf The engineering of Majorana zero modes in topological superconductors, a new paradigm for the realization of topological quantum computing and topology-based devices, has been hampered by the absence of materials with sufficiently large superconducting gaps. Recent experiments, however, have provided enthralling evidence for the existence of topological surface superconductivity in the iron-based superconductor \FST  possessing a full $s_\pm$-wave gap of a few meV. Here, we propose a mechanism for the emergence of topological superconductivity on the surface of \FST by demonstrating that the interplay between the $s_\pm$-wave symmetry of the superconducting gap, recently observed surface magnetism, and a Rashba spin-orbit interaction gives rise to several topological superconducting phases. Moreover, the proposed mechanism explains a series of experimentally observed hallmarks of topological superconductivity, such as the emergence of Majorana zero modes in the center of vortex cores and at the end of line defects, as well as of chiral Majorana edge modes along certain types of domain walls. We also propose that the spatial distribution of supercurrents near a domain wall is a characteristic signature measurable via a scanning superconducting quantum interference device that can distinguish between chiral Majorana edge modes and trivial in-gap states.  }

\noindent {\bf Introduction}\\
The non-Abelian braiding statistics of Majorana zero modes (MZMs) have opened a new route for the realization of topological quantum computing and topology-based devices \cite{Nayak2008}. Evidence for the existence of these modes has been observed in one- \cite{Mourik2012,NadjPerge2014,Das2012,Ruby2015,Pawlak2016,Kim2018,Manna2020} and two-dimensional (2D) \cite{Menard2017,Palacio-Morales2019,Menard2019} topological superconductors, however, their unambiguous identification has been experimentally hampered by the small superconducting gaps in these systems, which are often only of the order of a few hundred $\mu eV$. The recent report of topological superconductivity in the iron-based superconductor \FSTT, as evidenced by the observation of a surface Dirac cone \cite{Zhang2018,Rameau2019}, of MZMs in the vortex core \cite{Wang2018,Machida2019,Zhu2020} and at the end of line defects \cite{Chen2020}, and of chiral Majorana mode near domain walls \cite{Wang2020}, has therefore been greeted with much enthusiasm as this system possesses a significantly larger superconducting gap of a few meV. The origin of these observations was ascribed to \FST being a topological insulator whose surface Dirac cone is gapped out by proximity induced superconductivity, giving rise to topological surface superconductivity. However, the observation of a single $T_c$ that simultaneously gaps the Dirac cone and the (so-far assumed) topological trivial $\alpha$-, $\beta$- and $\gamma$-bands \cite{Zaki2019}, implies a coupling between them. This coupling, in turn, would destroy the topological character of the Dirac cone, raising the question as to the origin of the observed MZMs and of the underlying topological phase.

In this article, we propose that the origin of the observed Majorana modes lies in the emergence of topological superconductivity in the $\alpha$-, $\beta$- and $\gamma$-bands, arising from the interplay of a hard superconducting gap of $s_\pm$-wave symmetry, a Rashba spin-orbit interaction and surface magnetism. While a Rashba spin-orbit interaction occurs naturally at the surface of \FST due to the broken inversion symmetry, recent ARPES experiments \cite{Zaki2019} have provided strong evidence for the onset of surface magnetism at $T_c$. In particular, they observed that a gap opens up in the Dirac cone not only at the Fermi energy $E_F$, as expected from proximity induced superconductivity, but also at the Dirac point, which lies approximately 8 meV below $E_F$. The latter is a direct consequence of a broken time-reversal symmetry on the surface of \FSTT, with a non-vanishing magnetic moment perpendicular to the surface. We demonstrate that this interplay does not only give rise to robust topological superconducting phases on the surface of \FSTT, characterized by a $Z$ topological invariant, the Chern number, but also explains the various observations of Majorana zero modes in vortex cores \cite{Wang2018,Machida2019} and at the end of line defects \cite{Chen2020}, as well as the emergence of dispersive Majorana edge modes along domain walls \cite{Wang2020}.  As such, the existence of a Dirac cone within our scenario is secondary for the emergence of topological superconductivity and of the resulting MZMs, whose existence is primarily driven by the topological nature of the $\alpha$-, $\beta$- and $\gamma$-bands. In addition, we propose a novel experimental signature, the presence or absence of supercurrents along domain walls, which can distinguish topological Majorana modes from trivial in-gap states and can be imaged via a scanning superconducting quantum interference device (SQUID) \cite{Spanton2014}.\\

\noindent {\bf Results and Discussion}\\
 To investigate the emergence of topological superconductivity on the surface of \FSTT, we consider a two-dimensional 5-orbital model \cite{Gra09} extracted from an analysis of scanning tunneling spectroscopy (STS) experiments on the surface of clean \FST \cite{Sarkar2017}. In addition, we include in this model (a) the presence of surface magnetism, evidence for which was recently reported by ARPES experiments \cite{Zaki2019} through an exchange field, and (b) a Rashba spin orbit (RSO) interaction that arises from the breaking of the inversion symmetry on the surface [for detail, see supplementary materials (SM) Sec.~S1]. The resulting Hamiltonian in real space is given by
\begin{align}
 H_{0} =& -\sum_{a,b=1}^{5} \sum_{{\bf r},{\bf r'},\sigma} t^{ab}_{\bf r,r'} c_{{\bf r},a,\sigma}^{\dagger} c_{{\bf r'},b,\sigma} - \sum_{a=1}^{5} \sum_{{\bf r},,\sigma} \mu_{aa}  c_{{\bf r},a,\sigma}^{\dagger} c_{{\bf r},a,\sigma} \nonumber \\
 & +i \alpha \sum_{a=1}^{5} \sum_{{\bf r}, {\bm \delta}, \sigma, \sigma'} c_{{\bf r},a,\sigma}^\dag \left( {\bm \delta}
\times {\bm \sigma}\right)^z_{\sigma \sigma'} c^\pd_{{\bf r}+{\bm \delta},a, \sigma'}
\nonumber \\
& + J  \sum_{a=1}^{5} \sum_{{\bf r},\sigma, \sigma'} {\bf S}_{\bf r} \cdot c_{{\bf r},a, \sigma}^\dag {\boldsymbol \sigma}_{\sigma \sigma'} c^\pd_{{\bf r},a, \sigma'}  \nonumber \\
& + \sum_{a=1}^{5}  \sum_{\langle \langle {\bf r},{\bf r'} \rangle \rangle} \Delta^{aa}_{{\bf r}{\bf r}'} c_{{\bf r},a,\uparrow}^{\dagger} c_{{\bf r'},a,\downarrow}^{\dagger} + {\rm H.c.}
 \label{eq:Hamiltonian}
\end{align}
Here $a,b=1,...,5$ are the orbital indices corresponding to the $d_{xz}$-, $d_{yz}$-, $d_{x^2-y^2}$-, $d_{xy}$-, and $d_{3z^2-r^2}$-orbitals, respectively,  $-t^{ab}_{\bf rr'} $ represents the electronic hopping amplitude between orbital $a$ at site ${\bf r}$ and orbital $b$ at site ${\bf r'}$ on a two-dimensional square lattice, $\mu_{aa}$ is the on-site energy in orbital $a$, $c_{{\bf r},a,\sigma}^{\dagger} (c_{{\bf r},a,\sigma})$ creates (annihilates) an electron with spin $\sigma$ at site ${\bf r}$ in orbital $a$, and ${\bm \sigma}$ is the vector of spin Pauli matrices. The superconducting order parameter $\Delta^{aa}_{{\bf r}{\bf r}'}$ represents intra-orbital pairing between next-nearest neighbor Fe sites ${\bf r}$ and ${\bf r}'$ (in the 1 Fe unit cell), yielding a superconducting $s_\pm$-wave symmetry \cite{Sarkar2017}. Moreover, $\alpha$ denotes the Rashba spin-orbit interaction arising from the breaking of the inversion symmetry at the surface\,\cite{NadjPerge2014} with $\bm \delta$ being the vector connecting nearest neighbor sites. Due to the full superconducting gap, which suppresses Kondo screening, we consider the magnetic moments to be static in nature, such that ${\bf S}_{\bf r}$ is a classical vector representing the direction of a surface atom's spin located at ${\bf r}$, and $J$ is its exchange coupling with the conduction electron spin. The experimentally observed opening of a gap at the Dirac point \cite{Zaki2019} implies that a considerable fraction of the ordered magnetic moment is aligned perpendicular to the surface, such that for concreteness, we assume an out-of-plane ferromagnetic alignment. Within this model, topological superconductivity is thus confined to the surface of \FSTT. In the normal state, the above Hamiltonian yields three Fermi surfaces in the 1Fe Brillouin zone, two closed around the $\Gamma$-point, arising from the $\alpha$- and $\beta$-bands, and one closed around the $X/Y$-points, arising from the $\gamma$-band \cite{Sarkar2017}. Due to the orbital character of the three Fermi surfaces, the superconducting order parameter is only non-zero in the $d_{xz}$-, $d_{yz}$-, and $d_{xy}$-orbitals \cite{Sarkar2017}. The local density of states (LDOS) resulting form the above Hamiltonian in the superconducting state reproduces all salient features of the differential conductance, $dI/dV$, measured via STS (see SM Sec.~S1 and Fig.~S1), and in particular shows the existence of several superconducting gaps ranging from 1.6 meV to 2.4 meV.

We note that due to the particle-hole symmetry of the superconducting state, and the broken time-reversal symmetry arising from the presence of magnetic moments, \FST belongs to the topological class D \cite{Kitaev2009,Ryu2010}. For a two-dimensional system, the topological invariant is therefore given by the Chern number, which can be computed via \cite{Avron1983}
\begin{align}\label{eq:C}
 C & =  \frac{1}{2\pi i} \int_{\text{BZ}} d^2k \mathrm{Tr} ( P_{\bf{k}} [ \partial_{k_x} P_{\bf{k}}, \partial_{k_y} P_{\bf{k}} ] )  \nonumber \\
 P_{\bf{k}} & = \sum_{E_n(\bf{k}) < 0} |\Psi_n({\bf{k}}) \rangle \langle \Psi_n({\bf{k}})|
\end{align}
where $E_n({\bf{k}})$ and $|\Psi_n({\bf{k}}) \rangle$ are the eigenenergies and the eigenvectors of the Hamiltonian in Eq.(\ref{eq:Hamiltonian}), with $n$ being a band index, and the trace is taken over Nambu and spin space.\\

{\bf Topological Phase diagram}
\begin{figure}[t]
\centering
\includegraphics[width=8.5cm]{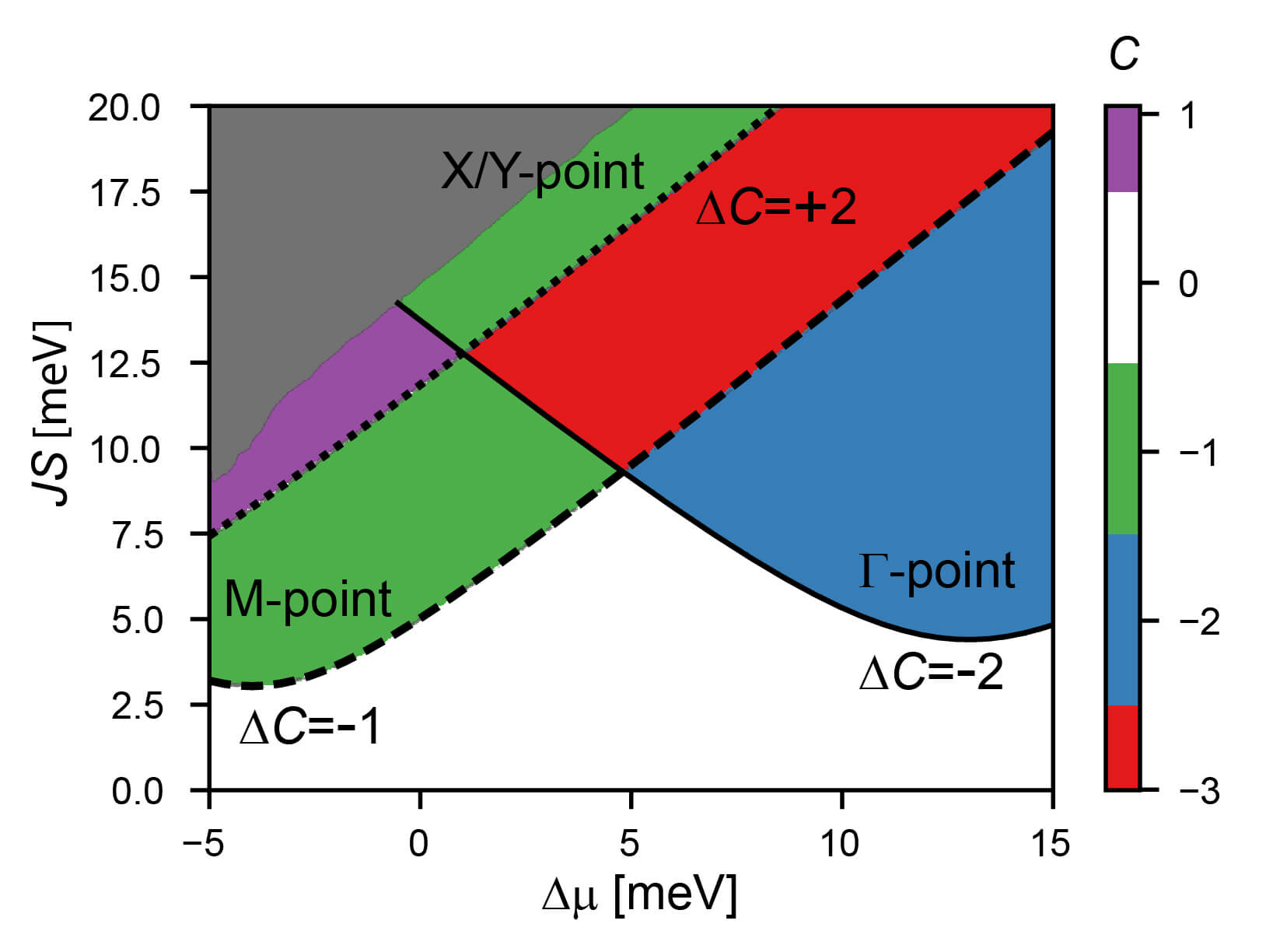}
\caption{{\bf  Topological phase diagram}. Topological phase diagram of FeSe$_{0.45}$Te$_{0.55}$ in the $(\Delta \mu, JS)$-plane with $\alpha=7$ meV. $\Delta \mu$ is a shift in the chemical potential in all orbitals from the value used in Eq.(\ref{eq:Hamiltonian}).
The solid, dashed and dotted black lines indicate gap closing at the $\Gamma$-, $X/Y$-, and $M$-points, respectively, which accompany the topological phase transitions. The gray area denotes the gapless region.
}
\label{fig:Fig1}
\end{figure}
The existence of a hard superconducting gap, a Rashba spin-orbit interaction and an out-of-plane ferromagnetic order are in general sufficient requirements for the emergence of topological surface superconductivity \cite{Ron2015,Li2016,Rachel2017,Crawford2020}. To demonstrate that this emergence is a robust phenomenon in \FST within the proposed model, we present in Fig.~\ref{fig:Fig1} the topological phase diagram -- in terms of the Chern number $C$ -- computed from Eq.(\ref{eq:C}) as a function of the effective magnetic exchange strength $JS$, with $S$ representing the ordered spin moment on the surface, and of a shift of the chemical potential, $\Delta \mu$, from its value extracted in Ref.\cite{Sarkar2017}. We note that already for rather weak magnetism, as reflected in a magnetic exchange coupling $JS$ of the order of a few meV, the system undergoes transitions into topological superconducting phases. An increase in the RSO interaction does not shift the position of the topological phase transitions, but increases the topological gap. While the presence of topological phases is robust against shifts in the chemical potential, varying $\Delta \mu$ induces transitions between topological phases characterized by different Chern numbers. With increasing $JS$ (keeping all other band parameters fixed), the Fermi surfaces eventually cross the nodal lines of the superconducting $s_\pm$-wave order parameter, and the system becomes gapless (as denoted by the gray area in Fig.~\ref{fig:Fig1}), and hence topologically trivial.  We note, however, that this gapless region of the phase diagram can be shifted to larger values of $JS$ by appropriately adjusting the band parameters in Eq.(\ref{eq:Hamiltonian}). The points in the $(\Delta \mu, JS)$-plane where the topological phase transitions occur are determined by the closing of the superconducting gap. Analytical expressions for these phase transition lines can be obtained (see SM Sec.~S2) and are plotted as solid, dashed and dotted lines in Fig.~\ref{fig:Fig1}) representing the closing of the gap at the $\Gamma$-, $X/Y$-, and $M$-points in the Brillouin zone, respectively. The $X/Y$-, and $M$-points possess a multiplicity of $m=2$ and $m=1$, leading to a change in the Chern number by $\Delta C = +2$ and $\Delta C = -1$, respectively. In contrast, while the $\Gamma$-point possesses a multiplicity of $m=1$, the band in which the gap closing occurs is two-fold degenerate, leading to a change in the Chern number by $\Delta C = -2$. Having established the presence of topological superconducting phases in \FSTT, we next turn to a discussion of their unique physical phenomena that have been observed experimentally.\\

{\bf MZM in a vortex core}
\begin{figure}[t]
\centering
\includegraphics[width=8.5cm]{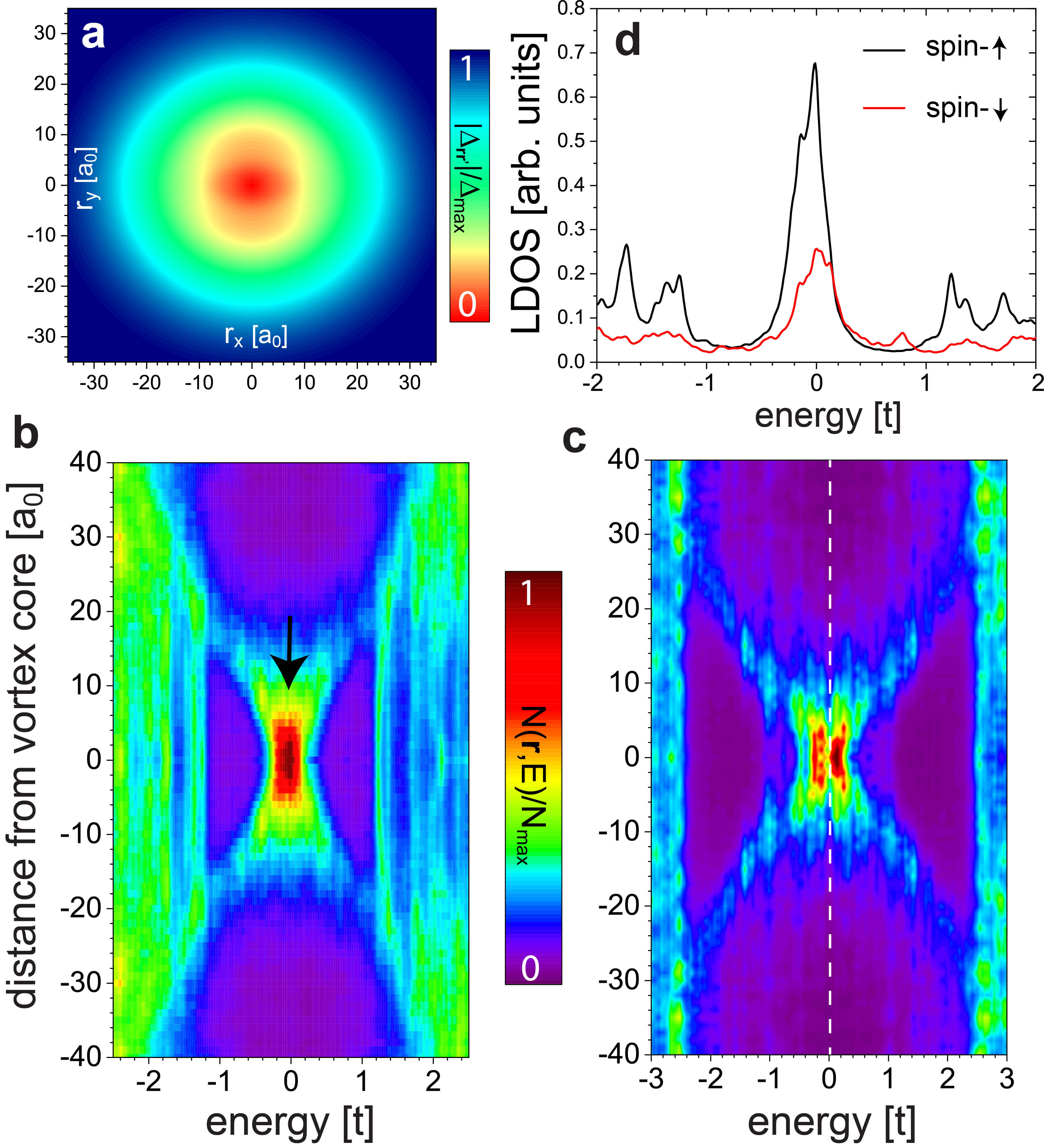}
\caption{{\bf MZM in a vortex core.} {\bf a} Spatial dependence of the superconducting order parameter in the $d_{xz}$-orbital near a vortex core for a magnetic field of $B=2$T. Linecut of the LDOS, $N({\bf r},E)$, through the center of the vortex core along the $x$-axis for {\bf b} the $C=-1$ topological phase with $(JS, \alpha, \Delta \mu)=(7.5, 7, 0)$ meV (the MZM is indicated by a black arrow) and {\bf c} the  topologically trivial phase ($C=0$) with $JS=\alpha=0$ (the dashed white line corresponds to $E=0$). {\bf d} Energy dependence of the spin-resolved LDOS at the vortex center in the topological $C=-1$ phase.}
\label{fig:Fig2}
\end{figure}
The experimental observation of MZMs localized in vortex cores \cite{Wang2018,Machida2019} represents a salient signature of the topological nature \cite{Fu2008} of the superconducting surface in \FSTT. As the detailed spatial, energy and spin-structure of these MZMs can provide important insight into the microscopic origin of the topological phase, we next investigate the electronic structure of MZMs near a vortex core in \FSTT. To this end, we implement the magnetic field via the Peierls substitution and compute the spatial dependence of the superconducting order parameters in the $d_{xz}$-, $d_{yz}$-, and $d_{xy}$-orbitals self-consistently (for details, see SM Sec.~S3). The resulting spatial structure of the superconducting order parameter in the $d_{xz}$-orbital, which vanishes at the center of the vortex, in the topological $C=-1$ phase is shown in Fig.~\ref{fig:Fig2}{\bf a} (the analogous plots for the $d_{yz}$-, and $d_{xy}$-orbitals are shown in SM Sec.~S3). We find that all three superconducting order parameters possess the same spatial symmetry as the orbitals they arise from, i.e., a $C_2$-symmetry in the $d_{xz}$-, and $d_{yz}$-orbitals, and a $C_4$-symmetry in the $d_{xy}$-orbital. To demonstrate the existence of a MZM localized at the vortex core, we present in Fig.~\ref{fig:Fig2}{\bf b} a linecut of the energy-resolved LDOS through the vortex which captures all salient features of the experimental observations: (i) the presence of a zero-energy state centered at the vortex core, (ii) a spatial extent of the MZM of about $15a_0$, and (iii) a decrease in the effective gap near the vortex core. In contrast, in the topologically trivial phase, the LDOS linecut [see Fig.~\ref{fig:Fig2}{\bf c}] does not show a zero-energy state, but only topologically trivial Caroli-de Gennes-Matricorn (CdGM) states \cite{Caroli1964}, similar to results obtained in other iron-based superconductors \cite{Uranga2016}. We therefore can identify the zero-energy state in the center of the vortex core shown in Fig.~\ref{fig:Fig2}{\bf b} as a MZM. We note, however, that though there is no MZM in the vortex core in the trivial phase [see Fig.~\ref{fig:Fig2}{\bf c}], the lowest energy trivial CdGM states are located at $E=\pm 0.05$ meV, and therefore might be difficult to resolve experimentally, and thus to distinguish from a true MZM.

An important clue as to the origin of the underlying topological phase arises form the spin-polarization of the MZM. In particular, the  spin- and energy-resolved LDOS at the center of the vortex core [see Fig.~\ref{fig:Fig2}{\bf d}] reveals a strong spin-polarization of the MZM, which is a direct consequence of the surface magnetism, and hence a characteristic signature of the here proposed origin of the topological superconductivity in \FSTT. In contrast, in previously proposed scenarios \cite{Wang2015,Zhang2018}, where the topological surface state arises from an interplay of a bulk topological insulator and superconductivity \cite{Zhang2018}, which does not break the time-reversal symmetry, no spin-polarization of the MZM is expected. We thus conclude that the scenario we propose for the emergence of topological surface superconductivity in \FST  does not only capture the salient experimental observations, but also possesses an essential signature in the strong spin-polarization of the vortex core MZM.\\

{\bf MZMs at the end of line defects}
\begin{figure}[t]
\centering
\includegraphics[width=8.5cm]{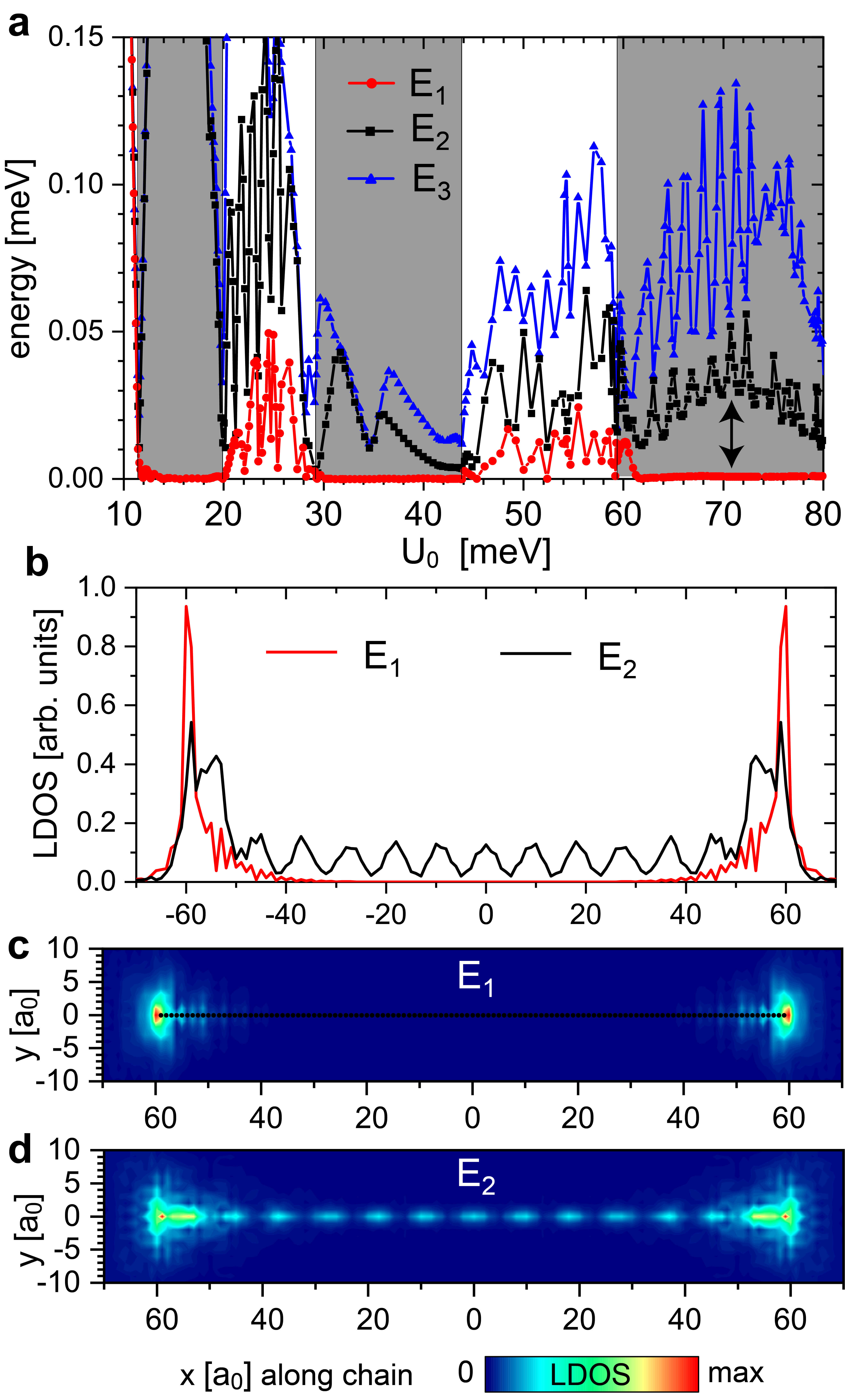}
\caption{{\bf MZMs and line defects.} {\bf a} Energies of the three lowest energy states, $E_1, E_2, E_3$ as a function of scattering strength $U_0$ for a line defect of length $L=119$ sites in the $C=-1$ phase with $(JS,\alpha,\Delta \mu)=(7.5, 7, 0)$meV. The regions of $U_0$ where the line defect is in a topological phase are shown using a gray background.{\bf b} Linecut of the spatial LDOS along the line defect for the two lowest energy states at $E_1, E_2$ and $U_0=71$ meV. Plot of the spatial LDOS at energies {\bf c} $E_1$ and {\bf d} $E_2$ for $U_0=71$ meV. Filled black circles in {\bf c} indicate the positions of the defects.}
\label{fig:Fig3}
\end{figure}
In addition to vortex cores, MZMs were also predicted to occur at the end of line defects that are embedded in topological $p_x + ip_y$-wave superconductors \cite{Wimmer2010}.
The recent observation of MZMs at the end of line defects on the surface of \FST \cite{Chen2020} has therefore raised the question of whether these MZMs are a direct signature of the underlying  topological superconducting phase, and thus represent a sufficient condition for its existence. To address this question, we represent the line defect for simplicity as a line of potential scatterers (though magnetic scatterers could also be realized \cite{Wu2010,Zhang2010}) described by the Hamiltonian
\begin{align}
H_{def} = U_0 \sum_{a=1}^{5} \sum_{{\bf R},\sigma} c_{{\bf R},a,\sigma}^{\dagger} c_{{\bf R},a,\sigma} \ ,
\end{align}
where $U_0$ is the potential scattering strength, and the sum runs over all sites ${\bf R}$ of the line defect In Fig.~\ref{fig:Fig3}{\bf a}, we present the energies of the three lowest energy states for a line defect of length $L=119$ in the $C=-1$ phase as a function of $U_0$. One can clearly discern three regions of $U_0$ where the lowest energy state is essentially located at zero energy (keeping in mind that the LDOS exhibits a full gap of 1.6 meV, see SM Sec.~1), suggesting the existence of a MZM. Further evidence arises from a line-cut of the LDOS along the line defect with $U_0=71$ meV, located in the rightmost gray region in Fig.~\ref{fig:Fig3}{\bf a}, for the lowest and second lowest energy states with energies $E_1$ and $E_2$, respectively [see Fig.~\ref{fig:Fig3}{\bf b}].
\begin{figure*}[t]
\centering
\includegraphics[width=17.5cm]{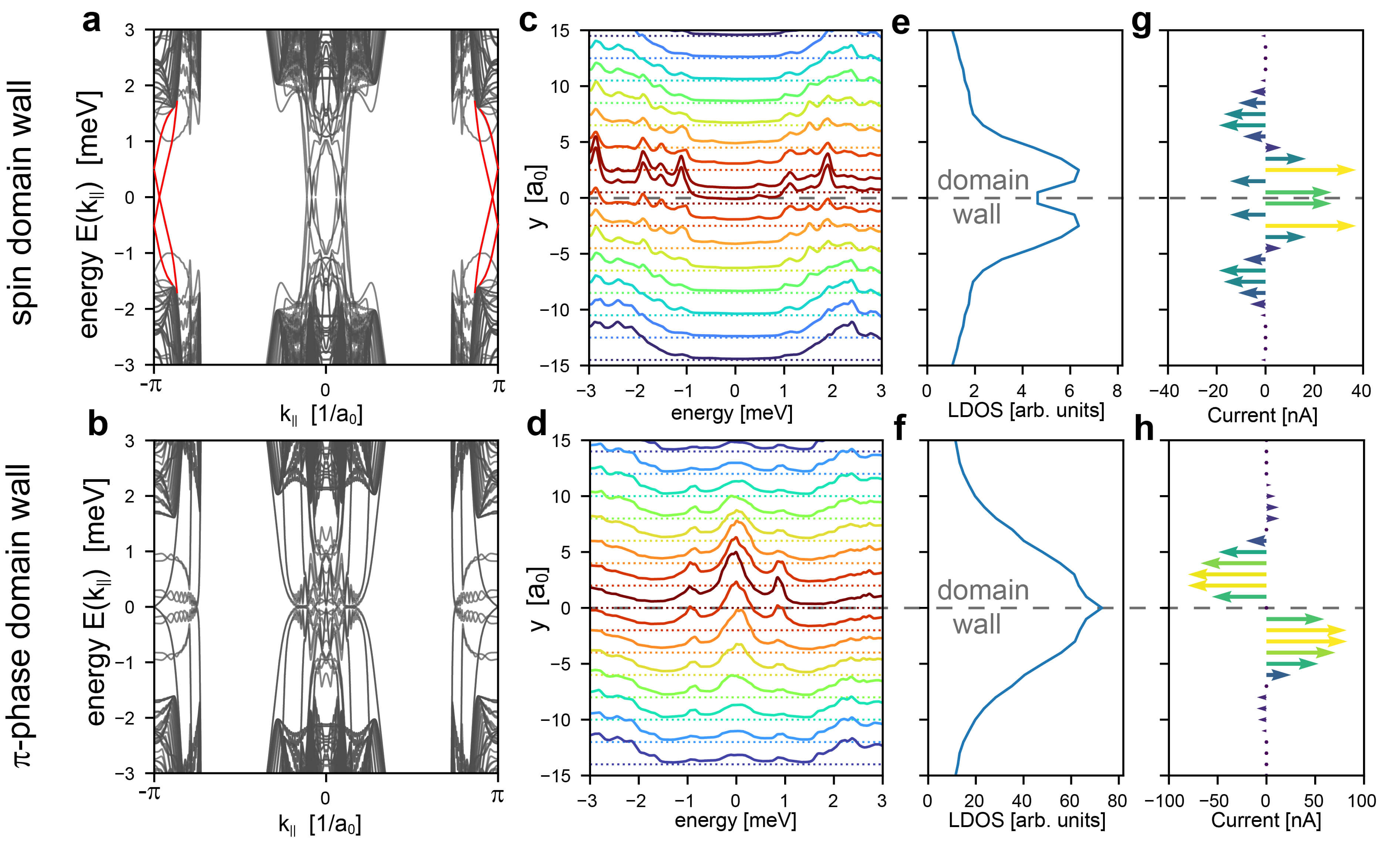}
\caption{{\bf Electronic structure near domain walls}. Upper (lower) row shows results for a spin domain wall (a $\pi$-phase domain wall). Electronic band structure as a function of momentum along the domain wall for {\bf a} a spin, and {\bf b} a $\pi$-phase domain wall.  Energy-resolved LDOS at {\bf c} the spin, and {\bf d} the $\pi$-phase domain wall along a linecut perpendicular to the domain wall. The position of the domain wall is indicated by a dashed gray line. Spatial plot of the zero-energy LDOS for {\bf e} a spin, and {\bf f} a $\pi$-phase domain wall. Spatial distribution of supercurrents near {\bf g} the spin, and {\bf h} the $\pi$-phase domain wall. For both types of domain walls, the system is in the $C=-1$ topological phase with $(JS, \alpha, \Delta \mu)=(7.5, 7, 0)$ meV.}
\label{fig:Fig4}
\end{figure*}
The spectral weight of the lowest energy state at $E_1$ is confined to the end of the line defect -- with essentially no spectral weight located inside the defect line -- reflecting the localized nature of the MZM. In contrast, the state at $E_2$ exhibits considerable spectral weight along the entire length of the line defect. This result is also confirmed by a spatial plot of the LDOS at energies $E_1$ and $E_2$, shown in Figs.~\ref{fig:Fig3}{\bf c} and {\bf d}, respectively. These results provide strong evidence for the existence of MZMs at the end of line defects in the topological $C=-1$ phase. The regions of $U_0$ where the line defect itself is in a topological phase and thus exhibits MZMs,   shown with a gray background in Fig.~\ref{fig:Fig3}{\bf a}, are bounded by topological phase transitions that are accompanied the closing of the superconducting gap. We note that while topological superconductivity on the surface of \FST is characterized by a $Z$ topological invariant, the line defect as a one-dimensional system possesses a topological $Z_2$ classification, similar to the Kitaev chain \cite{Bernevig2013}. Moreover, while we find similar results in the $C=-3$ phase (see SM Sec.~4), no zero-energy states exist at the end of a line defect when \FST is in the topological trivial ($C=0$) phase (see SM Sec.~4). We thus conclude that the existence of MZMs at the end of line defects is directly tied to and a sufficient condition for the existence of an underlying topological superconducting phase on the surface of \FSTT. However, the results shown in Fig.~\ref{fig:Fig3}{\bf a} also demonstrate that a failure to observe MZMs at line defects does not necessarily imply a trivial nature of the underlying superconducting phase. \\[0.5cm]

{\bf Chiral Majorana modes along domain walls} The bulk-boundary correspondence dictates that Majorana edge modes need to arise along domain walls that separate regions of different Chern numbers \cite{Bernevig2013}. Indeed, the observation of a nearly constant LDOS at a domain wall in \FST \cite{Wang2020} was recently interpreted as a signature of a chiral  Majorana mode. This raises the intriguing question not only as to which types of physical domain walls can give rise to the emergence of Majorana modes, but also of how to distinguish them from trivial, in-gap modes. To address this question, we calculate the electronic structure near two different types of domain walls: a spin domain wall at which the magnetic moment is inverted, i.e., ${\bf S} \rightarrow -{\bf S}$, and a $\pi$-phase domain wall, where the superconducting order parameter undergoes a $\pi$-phase shift, i.e., $\Delta \rightarrow -\Delta$, for all bands in \FSTT. As the spin domain wall separates regions with different Chern numbers (since ${\bf S} \rightarrow -{\bf S}$ implies $C \rightarrow -C$), the bulk boundary correspondence requires the emergence of dispersive Majorana edge modes that traverse the superconducting gap, as shown in Fig.~\ref{fig:Fig4}{\bf a} where we present the system's electronic band structure as a function of momentum $k_\parallel$ along the domain wall (Majorana modes are shown as red lines). In addition, the system also exhibits trivial in-gap modes, which do not connect the upper and lower bands. In contrast, regions separated by a $\pi$-phase domain wall exhibit the same Chern number, and the electronic band structure therefore only exhibits trivial in-gap states, as shown in Fig.~\ref{fig:Fig4}{\bf b}. In Figs.~\ref{fig:Fig4}{\bf c} and {\bf d}, we present the LDOS as a function of energy along a linecut perpendicular to the domain wall. In both cases, we find that the LDOS near the domain wall exhibits considerable spectral weight inside the superconducting gap, with the LDOS being nearly energy independent for the spin domain wall, but exhibiting a pronounced peak at zero-energy for the $\pi$-phase domain wall. A linecut of the zero-energy LDOS, shown in Figs.~\ref{fig:Fig4}{\bf e} and {\bf f}, however, reveals that the zero-energy state is localized close to the domain wall in both cases. Thus, the differences in the LDOS between these two types of domain walls is quantitative rather than qualitative in nature, and STS measurements might therefore not be able to distinguish between topological Majorana edge modes, and trivial in-gap states. However, a qualitative difference between these domain walls can be identified when considering the spatial structure of the induced supercurrents (see SM Sec.~S5). For a spin domain wall, the chirality of the induced supercurrent, which is determined by the sign of the Chern number, changes between the two separated regions, implying that the supercurrents associated with each region flow in the same direction along the domain wall \cite{Rachel2017}, as shown in Fig.~\ref{fig:Fig4}{\bf g}, yielding a non-vanishing net supercurrent. In contrast, for the $\pi$-phase domain wall, the chirality of the supercurrents in both regions is the same, implying that they flow in opposite directions along the domain wall [see Fig.~\ref{fig:Fig4}{\bf h}], yielding a vanishing net supercurrent. This qualitative difference, a non-zero net supercurrent for a spin domain wall and a vanishing net supercurrent for the $\pi$-phase domain wall, can be imaged using a SQUID, thus providing an unambiguous experimental signature to distinguish the existence of topological Majorana modes for a spin domain wall, from that of trivial in-gap states for a $\pi$-phase domain wall.\\

{\bf Conclusions} We have proposed a microscopic mechanism for the emergence of topological surface superconductivity in \FST arising from the interplay of surface magnetism, a Rashba spin-orbit interactions, and a hard superconducting gap with $s_\pm$-wave symmetry. This mechanism explains not only the emergence of robust topological phases already for weak surface magnetism, with effective magnetic exchange couplings of a few meV, but also the experimental observations of (i) MZMs in vortex cores, (ii) MZMs at the end of line defects, and (iii) chiral Majorana edge modes at domain walls. In addition, we demonstrated that by measuring supercurrents along domain walls using a SQUID it is possible to distinguish topological Majorana modes from trivial in-gap states.  Within our scenario, the existence of a Dirac cone is secondary for the emergence of topological superconductivity and the resulting Majorana modes, as they are primarily driven by the topological nature of the $\alpha$-, $\beta$- and $\gamma$-bands. One remaining interesting question for future work  pertains to the experimental signatures that might arise from the potential coupling between the topological $\alpha$-, $\beta$- and $\gamma$-bands and the Dirac cone.\\

\noindent{\bf Materials and Methods}\\
To compute the Chern number of \FSTT, yielding the topological phase diagram of Fig.~\ref{fig:Fig1}, we employ Eq.(\ref{eq:C}). To calculate the electronic structure of a vortex, we implement the magnetic field via the Peierls substitution and compute the spatial dependence of the superconducting order parameters in the $d_{xz}$-, $d_{yz}$-, and $d_{xy}$-orbitals self-consistently, as described in SM Sec.~S2.  The LDOS is computed by rewriting the Hamiltonian of Eq.(\ref{eq:Hamiltonian}) in terms of a Hamiltonian matrix, ${\hat H}$, in real and Nambu space, and calculating the retarded Greens function matrix ${\hat g}^{\rm r}$ using  ${\hat g}^{\rm r}(\omega) = \left[ (\omega + i \delta) {\hat 1} - {\hat H} \right]^{-1}$. The local, spin-resolved density of states, $N({\bf r},\omega, \sigma)$  at site ${\bf r}$ is then obtained via  $N({\bf r}, \sigma, \omega) = -\mathrm{Im}\left[\hat{g}^{\rm r}({\bf r}, \sigma; {\bf r}, \sigma; \omega)\right]/\pi$. The supercurrent is calculated using the Keldysh formalism, as described in SM Sec.~S5.\\

\noindent{\bf Acknowledgments}\\
The authors would like to thank A. Kreisel, C. Hess, and P.D. Johnson for stimulating discussions.\\

\noindent{\bf Funding}\\
 This work was supported by the U. S. Department of Energy, Office of Science, Basic Energy Sciences, under Award No. DE-FG02-05ER46225 (E.M., S.C., M.G. and D.K.M.) and through an ARC Future Fellowship (FT180100211) (S.R.).\\

\noindent{\bf Data and materials availability}: The authors declare that the main data supporting the findings of this study are available within the article and the supplementary materials. Extra data are available from the corresponding authors upon reasonable request.\\

\end{document}


\title{Origin of Topological Surface Superconductivity in \FST \\[0.25cm]
{\large Supplementary Material}}

\author{Eric Mascot$^{1}$, Sagen Cocklin$^{1}$, Martin Graham$^{1}$, Mahdi Mashkoori$^{2}$, Stephan Rachel,$^{2}$, and Dirk K. Morr$^{1}$}
\affiliation{$^{1}$University of Illinois at Chicago, Chicago, IL 60607, USA}
\affiliation{$^{2}$School of Physics, University of Melbourne, Parkville, VIC 3010, Australia}

\maketitle

\begin{center}
{\bf Section S1: Band parameters and local density of states in \FST}
\end{center}

\begin{figure}[h]
  \centering
\includegraphics[width=10cm]{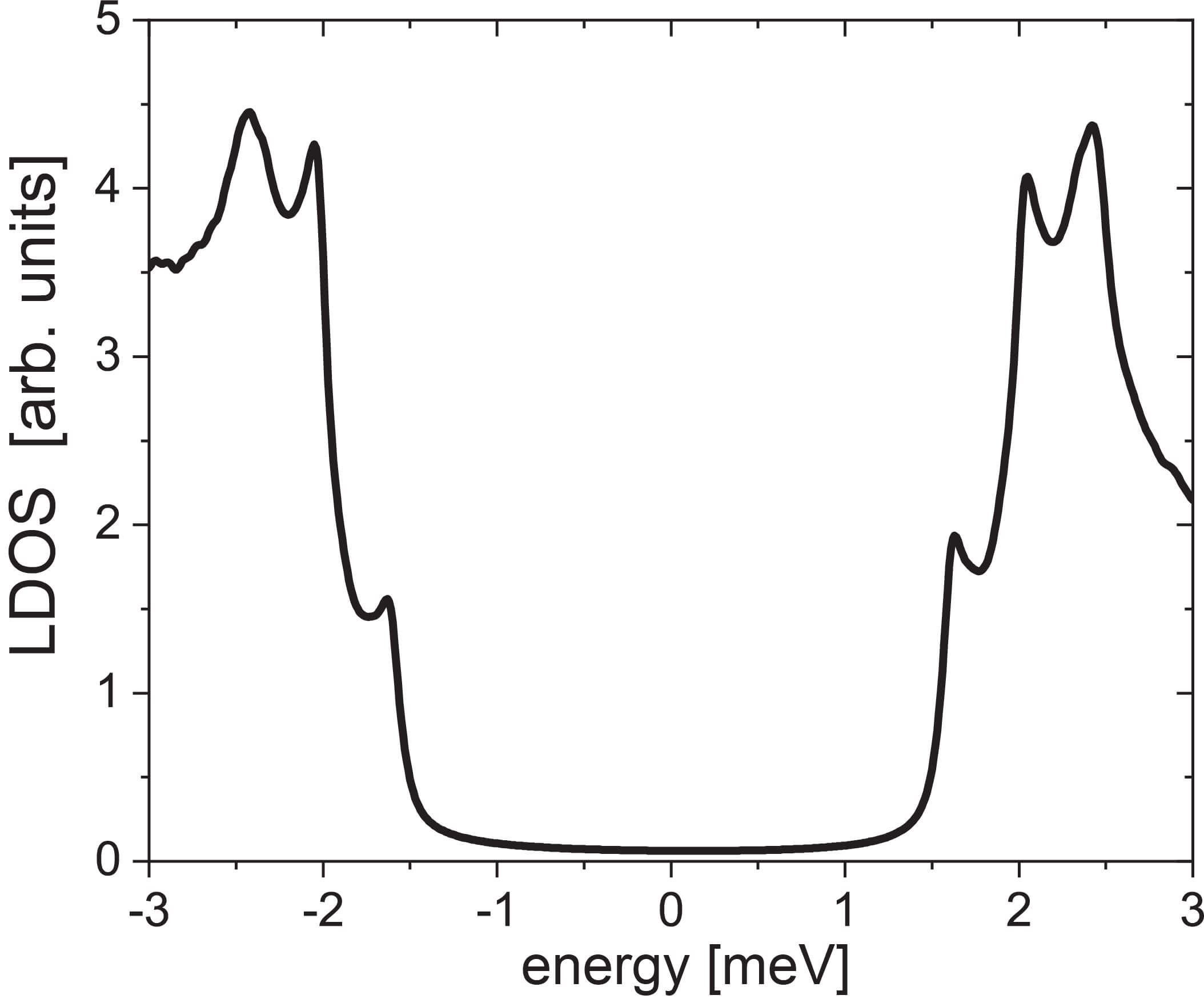}
\caption{LDOS of \FST in the topological $C=-1$ phase.}
  \label{fig:SI_Fig0}
\end{figure}

The band parameters in the Hamiltonian of Eq.(1) in the main text were extracted from scanning tunneling spectroscopy (STS) experiments on clean \FST \cite{Sarkar2017}. In order to reproduce a lineshape of the differential conductance $dI/dV$ measured in STS experiments in the topological $C=-1$ phase for $JS=7.5$ meV, $\alpha = 7$ meV, and $\Delta \mu = 0$,  we used the following intra-orbital superconducting order parameter reflecting pairing between next-nearest neighbor sites in the 1 Fe unit cell: $\Delta_{{\bf r}{\bf r}'}^{d_{xz}} = \Delta_{{\bf r}{\bf r}'}^{d_{yz}} = 1.1$ meV, $\Delta_{{\bf r}{\bf r}'}^{d_{xy}} = 0.76$ meV, and $\Delta_{{\bf r}{\bf r}'}^{d_{x^2-y^2}} = \Delta_{{\bf r}{\bf r}'}^{d_{3z^2-r^2}} = 0$. The resulting local density of states (LDOS), which is the sum of the LDOS for each orbital, is shown in Fig.~\ref{fig:SI_Fig0}.
It reproduces all salient feature of the experimental data: a hard superconducting gap, with the most pronounced coherence peaks located at $E = 2$ meV and 2.4 meV arising from the $\alpha$- and $\beta$-bands, and an additional in-gap peak around $E \approx 1.6$ meV arising from the $\gamma$-band. These band parameters are used for all figures in the main text that present results in the topological $C=-1$ phase. For the results shown in the topologically trivial phase in Fig.~2{\bf c} of the main text, we used $J=\alpha=\Delta \mu=0$ and $\Delta_{{\bf r}{\bf r}'}^{d_{xz}} = \Delta_{{\bf r}{\bf r}'}^{d_{yz}} = 0.55$ meV, $\Delta_{{\bf r}{\bf r}'}^{d_{xy}} = 0.$ 83meV, and $\Delta_{{\bf r}{\bf r}'}^{d_{x^2-y^2}} = \Delta_{{\bf r}{\bf r}'}^{d_{3z^2-r^2}} = 0$.\\[0.5cm]

\begin{center}
{\bf Section S2: Analytical expression for the topological phase transition lines}
\end{center}

An analytical expression for phase transition lines can be obtained by diagonalizing the Hamiltonian of Eq.(1) in the main text at time-reversal invariant high symmetry points in the 1Fe Brillouin zone. The closing of the superconducting gap at these points indicates a topological phase transition. One thus obtains the following expressions determining the relation between $JS$ and $\Delta \mu$ where a topological phase transition occurs and the gap closes (i) at the $\Gamma$-point [i.e., at ${\bf k}=(0,0)$]
\begin{align}
J &= \sqrt{ \left( E^{min}_{{\bf k}=(0,0)} - \Delta \mu \right)^2 + 4 \Delta_{{\bf r}{\bf r}'}^{d_{xz,yz}} } \ ,
\end{align}
with $E^{min}_{{\bf k}=(0,0)}=13$ meV is the energy of the lowest energy band in the normal state at ${\bf k}=(0,0)$, yielding the solid black line in Fig.~1 of the main text, (ii) at the $X/Y$-points [i.e., at ${\bf k} = (\pm \pi,0),(0,\pm \pi)$]
\begin{align}
J &=  \sqrt{\left( E^{min}_{{\bf k}=(0,\pi)} - \Delta \mu \right)^2 + 4 \Delta_{{\bf r}{\bf r}'}^{d_{xz,yz}}} \ ,
\end{align}
with $E^{min}_{{\bf k}=(0,\pi)} = -11$ meV is the energy of the lowest energy band in the normal state at ${\bf k}=(0,\pm \pi), (\pm \pi, 0)$, yielding the dashed black line in Fig.~1 of the main text, and (iii) at the $M$-points (i.e., at ${\bf k}=(\pm \pi, \pm \pi)$)
\begin{align}
J &= \sqrt{\left( E^{min}_{{\bf k}=(\pi,\pi)} - \Delta \mu \right)^2 + 4 \Delta_{{\bf r}{\bf r}'}^{d_{xy}} } \ ,
\end{align}
with $E^{min}_{{\bf k}=(0,\pi)} = -4$ meV is the energy of the lowest energy band in the normal state at ${\bf k}=(\pm \pi,\pm \pi)$, yielding the so dotted black line in Fig.~1 of the main text.\\[0.5cm]

\begin{center}
{\bf Section S3: MZM in a vortex core}
\end{center}

To calculate the structure of a vortex in the presence of a magnetic field, and in particular the spatial dependence of the superconducting order parameters, we employ the Peierls substitution \cite{Peierls1933,Hofstadter1976}, which replaces the electronic hopping terms according to
\begin{subequations}
\begin{eqnarray}
	t_{\vec{r}, \vec{r}'}^{ab}
	& \rightarrow &
	t_{\vec{r}, \vec{r}'}^{ab}
	e^{i\theta_{\vec{r}, \vec{r}'}}, \\
	i \alpha \left( {\bm \delta} \times {\bm \sigma}\right)^z_{\sigma \sigma'}
	& \rightarrow &
	i \alpha \left( {\bm \delta} \times {\bm \sigma}\right)^z_{\sigma \sigma'}
	e^{i\theta_{\vec{r}, \vec{r}+\vec{\delta}}}.
\end{eqnarray}
\end{subequations}
The Peierls phase factor is given by
\begin{equation}
	\theta_{\vec{r}, \vec{r}'}
	= \frac{\pi}{\phi_0} \int_{\vec{r}}^{\vec{r}'} \vec{A}(\vec{s}) \cdot d\vec{s},
\end{equation}
where $\phi_0 = h/2e$ is the superconducting flux quantum and $\vec{A}$ is the magnetic vector potential. Here, we use the symmetric gauge $\vec{A}(\vec{r}) = \frac{1}{2} B \hat{z} \times \vec{r}$.
\begin{figure}[h]
  \centering
\includegraphics[width=12cm]{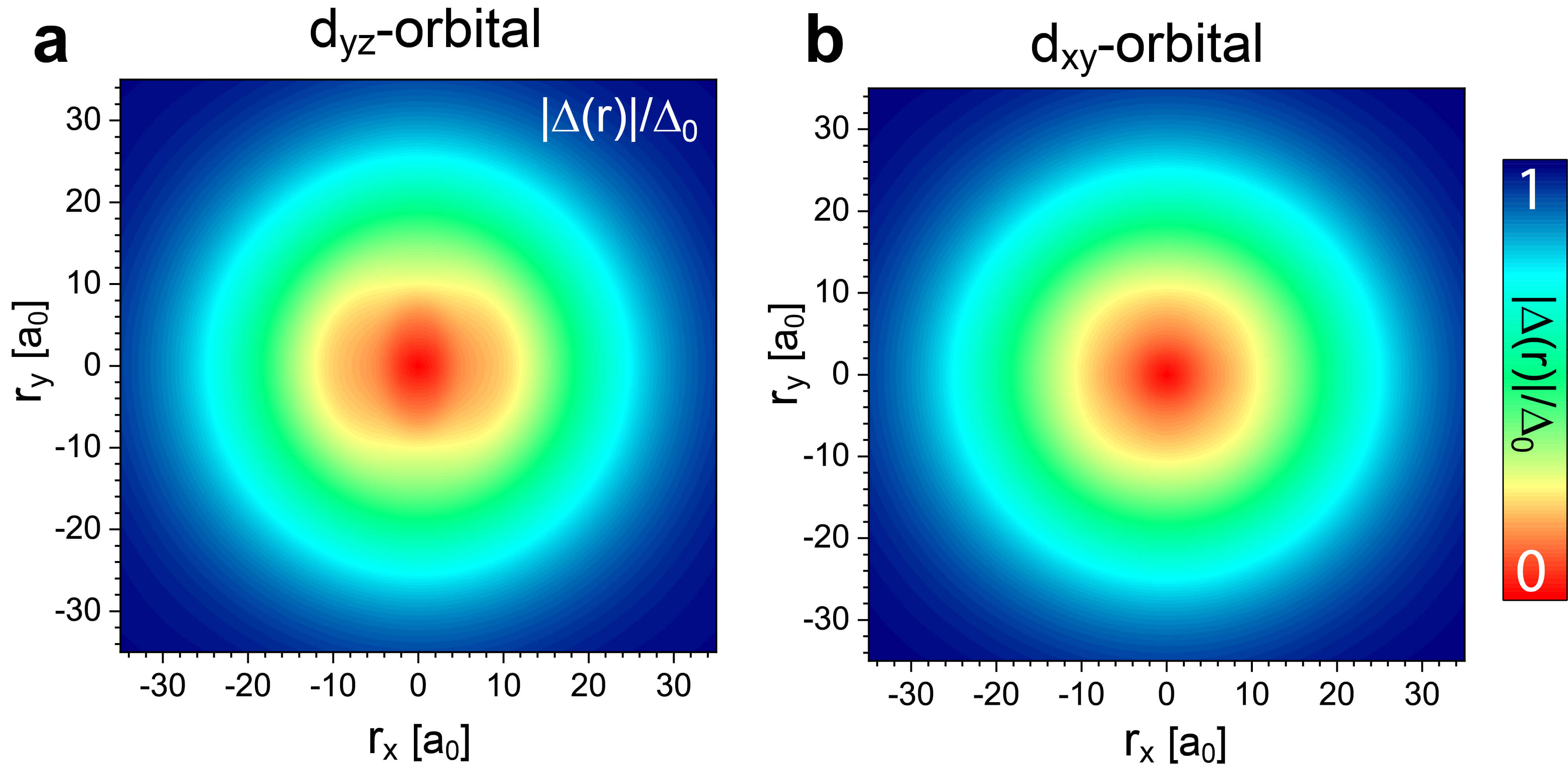}
\caption{Spatial plot of the superconducting order parameter in the {\bf a} $d_{yz}$-, and {\bf b} $d_{xy}$-orbital in the $C=-1$ phase with $( JS, \alpha, \Delta \mu) = (7.5, 7, 0)$ meV.}
  \label{fig:SI_Fig1}
\end{figure}
The superconducting order parameters in the presence of the magnetic field are self-consistently computed using the gap equation,
\begin{equation}
	\Delta_{\vec{r}, \vec{r}'}^{aa}
	= V_{\vec{r}, \vec{r}'}^{aa} \braket{c_{\vec{r}', a, \dw} c_{\vec{r}, a, \up}}.
\label{eq:gap}
\end{equation}
The pairing interaction, $V_{\vec{r}, \vec{r}'}^{aa}$, is chosen such that it reproduces the values of the superconducting order parameters $\Delta_{0}^{d_{xz}} = \Delta_{0}^{d_{yz}} = 1.1 \; \rm{meV}$, $\Delta_{0}^{d_{xy}} = 0.76 \; \rm{meV}$, and $\Delta_{0}^{d_{x^2-y^2}} = \Delta_{0}^{d_{3z^2-r^2}} = 0 \; \rm{meV}$ for a clean system (i.e., in the absence of a magnetic field), as discussed in Sec.~S1. This yields the following values
\begin{subequations}
\begin{align}
& V_{\vec{r}, \vec{r}'}^{d_{xz}}  = V_{\vec{r}, \vec{r}'}^{d_{yz}} =  62.8 \; \rm{meV} \\
& V_{\vec{r}, \vec{r}'}^{d_{xy}}  = 54.1\; \rm{meV}\\
& V_{\vec{r}, \vec{r}'}^{d_{x^2-y^2}} = V_{\vec{r}, \vec{r}'}^{d_{3z^2-r^2}} = 0
\end{align}
\end{subequations}
The expectation values on the right hand side of Eq.(\ref{eq:gap}) are calculated using the kernel polynomial method \cite{Weisse2006,Nagai2012} with 200 moments and the gap equation was solved iteratively using Anderson's method \cite{Anderson1965,Johnson1988} until the largest error was below $8.5 \; \rm{\mu eV}$.
\begin{figure}[h]
  \centering
\includegraphics[width=15cm]{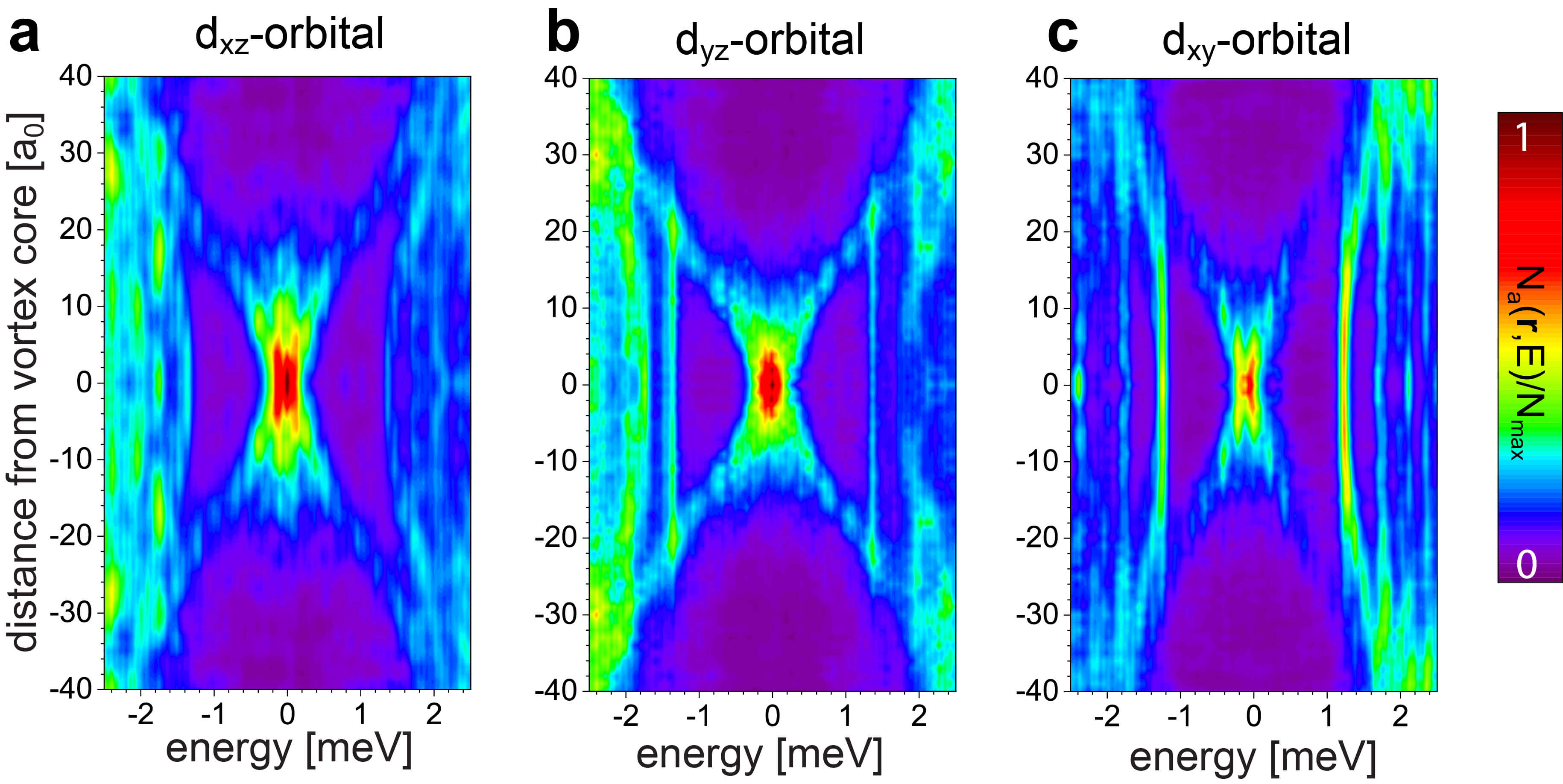}
\caption{Linecut of the orbitally-resolved LDOS, $N({\bf r},E)$, through the center of the vortex core along the $x$-axis in the {\bf a} $d_{xz}$-, {\bf b} $d_{yz}$-, and {\bf c} $d_{xy}$-orbitals for the $C=-1$ topological phase with $( JS, \alpha, \Delta \mu) = (7.5, 7, 0)$ meV. The MZM localized in the vortex core is clearly visible in all three orbitals. }
  \label{fig:SI_Fig2}
\end{figure}
In Fig.~\ref{fig:SI_Fig1} we plot the resulting spatial structure of the superconducting order parameters in the $d_{yz}$- [Fig.~\ref{fig:SI_Fig1}{\bf a}] and $d_{xy}$-orbitals  [Fig.~\ref{fig:SI_Fig1}{\bf b}] [that in the $d_{xz}$-orbital is shown in Fig.~2{\bf a} in the main text] in the $C=-1$ phase; they reveal the presence of a vortex core where the superconducting order parameter vanishes. For all three orbitals, the spatial symmetry of the superconducting order parameter reflects that of the underlying orbital.

In Fig.~\ref{fig:SI_Fig2} we present a linecut of the orbitally-resolved LDOS, $N_a({\bf r},E)$, through the center of the vortex core for the $d_{xz}$- [Fig.~\ref{fig:SI_Fig2}{\bf a}], $d_{yz}$- [Fig.~\ref{fig:SI_Fig2}{\bf b}], and $d_{xy}$-orbitals [Fig.~\ref{fig:SI_Fig2}{\bf c}] in the $C=-1$ topological phase. The MZM localized in the vortex core is clearly visible not only in the total LDOS presented in Fig.~2{\bf b} of the main text, but also in the LDOS of each orbital.\\[0.5cm]

\begin{center}
{\bf Section S4: Electronic structure of a line of defects}
\end{center}

We showed in the main text, that when \FST is in the $C=-1$ phase, a line defect can exhibit MZMs at its ends for certain values of the scattering potential $U_0$.
\begin{figure}[h]
  \centering
\includegraphics[width=15cm]{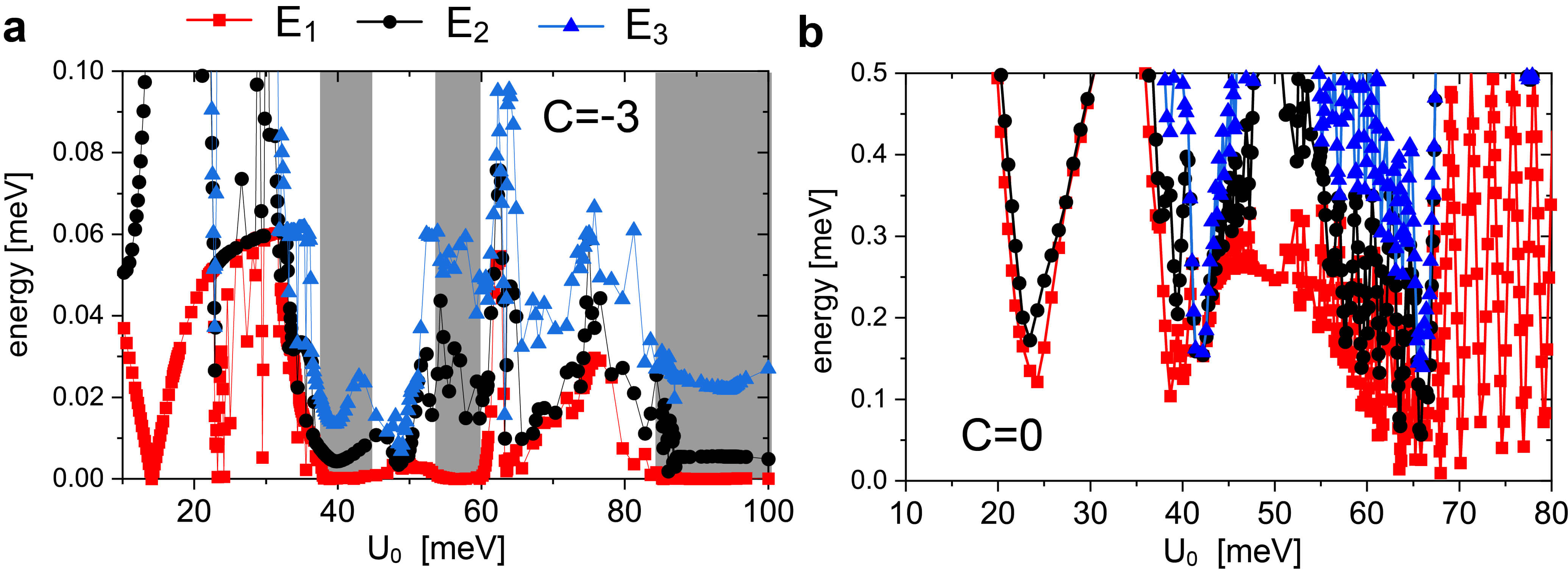}
\caption{Energies of the three lowest energy states, $E_1, E_2, E_3$ as a function of scattering strength $U_0$ for a line defect with $L=119$ sites in {\bf a} the topological $C=-3$ phase with $(JS,\alpha,\Delta \mu)=(12, 7, 5)$ meV, and $\Delta_{{\bf r}{\bf r}'}^{d_{xz}} = \Delta_{{\bf r}{\bf r}'}^{d_{yz}} = 2.2$ meV, $\Delta_{{\bf r}{\bf r}'}^{d_{xy}} = 1.52$ meV, and {\bf b} the topological trivial $C=0$ phase with $(JS, \alpha, \Delta \mu)=(0,0,0)$, and  $\Delta_{{\bf r}{\bf r}'}^{d_{xz}} = \Delta_{{\bf r}{\bf r}'}^{d_{yz}} = 1.1$ meV, $\Delta_{{\bf r}{\bf r}'}^{d_{xy}} = 0.76$ meV.  The regions of $U_0$ where MZMs exist are shown using a gray background in {\bf a}.}
  \label{fig:SI_Fig3}
\end{figure}
A similar result is also obtained in the $C=-3$ phase as shown in Fig.~\ref{fig:SI_Fig3}{\bf a} where we present the three lowest energy states a function of scattering strength $U_0$ for a line defect with $L=119$ sites. While MZMs also exist in the $C=-3$ phase (the regions of $U_0$ where MZMs exist are shown using a gray background), they do so over a smaller range of $U_0$ than in the $C=-1$ phase. We note that while topological superconductivity on the two-dimensional surface of \FST is characterized by a $Z$ topological invariant, the Chern number, the line defect as a one-dimensional system possesses a topological $Z_2$ classification, similar to the Kitaev chain. As a result, the line defect possesses a single set of MZMs (one at each end) when it is in the topologically non-trivial phase, independent of the Chern number of the underlying topological phase of \FSTT.

In contrast, in the topologically trivial $C=0$ phase, the line defect does not possess any MZMs, as follows from a plot of the three lowest energy states in Fig.~\ref{fig:SI_Fig3}{\bf b} as a function of $U_0$. These results support our conclusion that the existence of MZM at the end of line defects is a sufficient condition for the existence of an underlying topological superconducting phase on the surface of \FSTT.\\[0.5cm]

\begin{center}
{\bf Section 5: Persistent supercurrents along domain walls}
\end{center}

The persistent supercurrent associated with the hopping of an electron from a site ${\bf r}$ to another site ${\bf r}+{{\bm \delta}}$ can be computed via
\begin{align}
 I_{{\bf r},  {\bf r}+{ {\bm \delta}}} &=
 -\frac{2e}{\hbar} \sum_{\sigma, \sigma'} \sum_{a, b = 1}^5 \int \frac{d\omega}{2\pi}
  \text{Re} \left[
   \left(
    -t^{ab}_{{\bf r},  {\bf r}+{ {\bm \delta}}} \delta_{\sigma \sigma'}
    + i \alpha \left( {\bm \delta} \times {\bm \sigma}\right)^z_{\sigma \sigma'}
   \right)
   g^{<}_{b,\sigma';a,\sigma}({\bf r}+{ {\bm \delta}}, {\bf r})
  \right] \ ,
 \label{eq:I1}
\end{align}
if ${\bf r}$ and ${\bf r}+{{\bm \delta}}$ are nearest neighbor sites, and via
\begin{align}
 I_{{\bf r},  {\bf r}+{ {\bm \delta}}} &=
 -\frac{2e}{\hbar} \sum_{\sigma, \sigma'} \sum_{a, b = 1}^5 \int \frac{d\omega}{2\pi}
  \text{Re} \left[
   -t^{ab}_{{\bf r},  {\bf r}+{ {\bm \delta}}} \delta_{\sigma \sigma'}
   g^{<}_{\sigma',b;\sigma,a}({\bf r}+{ {\bm \delta}}, {\bf r})
  \right]
 \ ,
 \label{eq:I2}
\end{align}
otherwise, where $-t^{ab}_{{\bf r},  {\bf r}+{ {\bm \delta}}}$ is the hopping amplitude and $\alpha$ is the Rashba spin-orbit interaction given in Eq.(1) of the main text, and
${g}^{r,a,<}_{\sigma,a;\sigma',b}({\bf r},{\bf r}+{ {\bm \delta}},\omega)$ are the $(\sigma,a;\sigma',b)$ elements in Nambu space of the retarded, advanced, or lesser Green's function matrices. The Green's function matrix in Matsubara time is defined via
\begin{align}
  {\hat g}({\bf r},{\bf r}+{{\bm \delta}},\tau) & = -\langle T_\tau \Psi_{{\bf r}}(\tau) \Psi_{{\bf r}+{ {\bm \delta}}}^\dagger (0) \rangle
  \label{eq:gf}
\end{align}
where the spinors are defined via
\begin{subequations}
\begin{align}
 \Psi_{{\bf r}}^\dagger &= \left (
  \psi_{{\bf r},\uparrow}^\dagger,
  \psi_{{\bf r},\downarrow}^\dagger,
  \psi_{{\bf r},\uparrow}^T,
  \psi_{{\bf r},\downarrow}^T
 \right ) \\
 \psi_{{\bf r},\sigma}^\dagger &= \left(
  c^\dagger_{{\bf r},\sigma,xz},
  c^\dagger_{{\bf r},\sigma,yz},
  c^\dagger_{{\bf r},\sigma,x^2-y^2},
  c^\dagger_{{\bf r},\sigma,xy},
  c^\dagger_{{\bf r},\sigma,3z^2-r^2}
 \right) \ .
\label{eq:spinor}
\end{align}
\end{subequations}
To obtain the above Greens functions for the system, we diagonalize the real space Hamiltonian in Eq.(1) of the main text, yielding energy eigenvalues $E_k$ and eigenvectors $u_{\sigma,a;k}({\bf r})$. The Greens functions can then be computed using
\begin{subequations}
\begin{align}
g^{r}_{\sigma,a;\sigma',b}(\mathbf{r}_s,\mathbf{r}^\prime_s,\omega) &= \sum_{k}  \frac{u_{\sigma,a;k}(\mathbf{r}_s) u_{\sigma',b;k}^*(\mathbf{r}^\prime_s)}{\omega - E_k + i\delta} \\
g^{a}_{\sigma,a;\sigma',b} (\mathbf{r}_s,\mathbf{r}^\prime_s,\omega) &= \sum_{k}\frac{u_{\sigma,a;k}(\mathbf{r}_s) u_{\sigma',b;k}^*(\mathbf{r}^\prime_s)}{\omega - E_k - i\delta} \\
g^{<}_{\sigma,a;\sigma',b} (\mathbf{r}_s,\mathbf{r}^\prime_s,\omega) &= - n_F(\omega)\sum_{k} u_{\sigma,a;k}(\mathbf{r}_s) u_{\sigma',b;k}^*(\mathbf{r}^\prime_s) \left ( \frac{1}{\omega - E_k + i\delta} - \frac{1}{\omega - E_k - i\delta}  \right ) \nonumber \nonumber \\
&=  2 \pi i n_F(\omega)\sum_{k} u_{\sigma,a;k}(\mathbf{r}_s) u_{\sigma',b;k}^*(\mathbf{r}^\prime_s)  \delta({\omega - E_k}) \ .
\label{eq:GFs}
\end{align}
\end{subequations}
Note that the energy eigenvalues come in pairs $\pm E_k$, and that the summation in the above equation runs over all eigenvalues.
Since the Hamiltonian in Eq.(1) of the main text contains hopping terms between up to 5th neighbor sites \cite{Sarkar2017}, plotting all currents becomes prohibitive. In the plots presented in Figs.~4{\bf g} and {\bf h} of the main text, we therefore plot for clarity the directed sum of supercurrents, defined via
\begin{align}
 I_{{\bf r}} &= \sum_{{\bm \delta}}  I_{{\bf r}, {\bf r}+{ {\bm \delta}}} \hat{\bm \delta} \ ,
\end{align}
where $\hat{\bm \delta} = {\bm \delta}/|{\bm \delta}|$ is the unit vector in the direction of ${\bm \delta}$, and the sum runs over all sites ${\bf r}+{ {\bm \delta}}$
for which $-t^{ab}_{{\bf r},  {\bf r}+{ {\bm \delta}}} \not = 0$.
